\documentclass{elsart}

\usepackage{amssymb}
\usepackage{graphicx}
\begin{document}
%
%
\def\be{\begin{equation}}
\def\ee{\end{equation}}
\def\bea{\begin{eqnarray}}
\def\eea{\end{eqnarray}}
\def\beq{\begin{equation}}
\def\eeq{\end{equation}}
\def\bq{\begin{quote}}
\def\eq{\end{quote}}
\def\gappeq{\mathrel{\rlap {\raise.5ex\hbox{$>$}} {\lower.5ex\hbox{$\sim$}}}}
\def\lappeq{\mathrel{\rlap{\raise.5ex\hbox{$<$}} {\lower.5ex\hbox{$\sim$}}}}
\def\PR{{\it Phys.~Rev.~}}
\def\PRL{{\it Phys.~Rev.~Lett.~}}
\def\NP{{\it Nucl.~Phys.~}}
\def\NPBPS{{\it Nucl.~Phys.~B (Proc.~Suppl.)~}}
\def\PL{{\it Phys.~Lett.~}}
\def\PRep{{\it Phys.~Rep.~}}
\def\AP{{\it Ann.~Phys.~}}
\def\CMP{{\it Comm.~Math.~Phys.~}}
\def\JMP{{\it J.~Math.~Phys.~}}
\def\NC{{\it Nuovo~Cim.~}}
\def\SJNP{{\it Sov.~J.~Nucl.~Phys.~}}
\def\SPJETP{{\it Sov.~Phys.~J.E.T.P.~}}
\def\ZP{{\it Zeit.~Phys.~}}
\def\JP{{\it Jour.~Phys.~}}
\def\JHEP{{\it J.~High~En.~Phys.~}}
\def\vol#1{{\bf #1}}
\def\vyp#1#2#3{\vol{#1} (#2) #3}
\def\Im{\,\hbox{Im}\,}
\def\Re{\,\hbox{Re}\,}
\def\tr{\,{\hbox{tr}}\,}
\def\Tr{\,{\hbox{Tr}}\,}
\def\det{\,{\hbox{det}}\,}
\def\Det{\,{\hbox{Det}}\,}
\def\neath#1#2{\mathrel{\mathop{#1}\limits_{#2}}}
\def\ker{\,{\hbox{ker}}\,}
\def\dim{\,{\hbox{dim}}\,}
\def\ind{\,{\hbox{ind}}\,}
\def\sgn{\,{\hbox{sgn}}\,}
\def\mod{\,{\hbox{mod}}\,}
\def\apm#1{\hbox{$\pm #1$}}
\def\epm#1#2{\hbox{${\lower1pt\hbox{$\scriptstyle +#1$}}
\atop {\raise1pt\hbox{$\scriptstyle -#2$}}$}}
\def\neath#1#2{\mathrel{\mathop{#1}\limits_{#2}}}
\def\gsim{\mathrel{\rlap{\lower4pt\hbox{\hskip1pt$\sim$}}
    \raise1pt\hbox{$>$}}}         
\def\eg{{\it e.g.}}
\def\ie{{\it i.e.}}
\def\viz{{\it viz.}}
\def\etal{{\it et al.}}
\def\rhs{right hand side}
\def\lhs{left hand side}
\def\toinf#1{\mathrel{\mathop{\sim}\limits_{\scriptscriptstyle
{#1\rightarrow\infty }}}}
\def\tozero#1{\mathrel{\mathop{\sim}\limits_{\scriptscriptstyle
{#1\rightarrow0 }}}}
\def\frac#1#2{{{#1}\over {#2}}}
\def\half{\hbox{${1\over 2}$}}\def\third{\hbox{${1\over 3}$}}
\def\quarter{\hbox{${1\over 4}$}}
\def\smallfrac#1#2{\hbox{${{#1}\over {#2}}$}}
\def\pbp{\bar{\psi }\psi }
\def\vevpbp{\langle 0|\pbp |0\rangle }
\def\as{\alpha_s}
\def\tr{{\rm tr}}\def\Tr{{\rm Tr}}
\def\eV{{\rm eV}}\def\keV{{\rm keV}}
\def\MeV{{\rm MeV}}\def\GeV{{\rm GeV}}\def\TeV{{\rm TeV}}
\def\MS{\hbox{$\overline{\rm MS}$}}
\def\blackbox{\vrule height7pt width5pt depth2pt}
\def\matele#1#2#3{\langle {#1} \vert {#2} \vert {#3} \rangle }
\def\VertL{\Vert_{\Lambda}}\def\VertR{\Vert_{\Lambda_R}}
\def\Real{\Re e}\def\Imag{\Im m}
\def\bp{\bar{p}}\def\bq{\bar{q}}\def\br{\bar{r}}
\catcode`@=11 
\def\slash#1{\mathord{\mathpalette\c@ncel#1}}
 \def\c@ncel#1#2{\ooalign{$\hfil#1\mkern1mu/\hfil$\crcr$#1#2$}}
\def\lsim{\mathrel{\mathpalette\@versim<}}
\def\gsim{\mathrel{\mathpalette\@versim>}}
 \def\@versim#1#2{\lower0.2ex\vbox{\baselineskip\z@skip\lineskip\z@skip
       \lineskiplimit\z@\ialign{$\m@th#1\hfil##$\crcr#2\crcr\sim\crcr}}}
\catcode`@=12 
\def\twiddles#1{\mathrel{\mathop{\sim}\limits_
                        {\scriptscriptstyle {#1\rightarrow \infty }}}}
\def\sitontop#1#2{\mathrel{\mathop{\scriptstyle #1}\limits_{\scriptstyle #2}}}
\newcommand{\stw}{\mbox{$\sin^2\theta_W$}}
\newcommand{\nub}{\overline{\nu}}
\newcommand{\nue}{\nu_{e}}
\newcommand{\numu}{\nu_{\mu}}
\newcommand{\nubmu}{\overline{\nu_{\mu}}}
\newcommand{\nube}{\overline{\nu_{e}}}
\def\tr{\,{\hbox{\rm tr}}\,}
\def\epm#1#2{\hbox{${\lower1pt\hbox{$\scriptstyle +#1$}}
\atop {\raise1pt\hbox{$\scriptstyle -#2$}}$}}
\def\wup{{W^+}}
\def\wum{{W^-}}
\def\wupm{{W^\pm}}
\def\wump{{W^\mp}}
\def    \ie             {{\em i.e.\/} } 
\def    \etal             {{\em et al.\/} } 
\def    \eg             {{\em e.g.\/} } 
\def\MS{\hbox{$\overline{\rm MS}$}}
\def    \lsim {\raisebox{-3pt}{$\>\stackrel{<}{\scriptstyle\sim}\>$}}  
\def    \gsim {\raisebox{-3pt}{$\>\stackrel{>}{\scriptstyle\sim}\>$}}  
\def    \gtrsim {\raisebox{-3pt}{$\>\stackrel{>}{\scriptstyle\sim}\>$}}  
\def    \esim {\raisebox{-3pt}{$\>\stackrel{-}{\scriptstyle\sim}\>$}}  
\def\wup{{W^+}}
\def\wum{{W^-}}
\def\wupm{{W^\pm}}
\def\wump{{W^\mp}}

\begin{frontmatter}


\setcounter{page}{0}
\begin{flushright}
{\tt hep-ph/0109219}\\
{RM3-TH 01/11}\\
\end{flushright}
\vglue.3cm

\title{
Highlights of Short Baseline Physics\\ at  a  Neutrino Factory}

\author{Stefano Forte\thanksref{leave}}

\thanks[leave]{On leave from INFN, Sezione di Torino, Italy}

\address{INFN, Sezione di Roma III, via della Vasca Navale 84,
  I--00146 Roma, Italy}
\vglue2.cm

\begin{abstract}
We review the physics potential for experiments with intense neutrino
beams at the front--end of a muon storage ring, stressing  the way the
spin and flavour structure of neutrino interactions with matter can be
used to shed light on the structure of the
strong and electroweak
interactions. 
Specifically, we discuss precision tests of the 
standard model, studies of polarized and unpolarized 
structure functions of the nucleon,  and several new and exotic physics items.
\end{abstract}


\end{frontmatter}
\vskip 2.5cm
\begin{center}
 Invited plenary talk at
{\bf NuFact 01}\\
Tsukuba, Japan, May 2001\\
{\it to be published in the proceedings}
\end{center}

\newpage
\section{Probing Matter with Intense Neutrino Beams}

Front--end physics at a neutrino factory is based on the realization
that, because of the flavour and spin structure of the coupling of
neutrinos to weak currents,  a neutrino beam is a unique probe of the structure
of the standard model and of the structure of the nucleon. A neutrino
beam thus has a  greater physics potential than conventional electron
or muon beams, provided the intensity of the beam is high enough.
An accurate assessment of the physics potential of the experiments
which could be performed with an intense neutrino beam cannot abstract
from the fact that the time scale for the construction of a neutrino
factory is of order of ten years: clearly, it is difficult to envisage
what the `standard model' will be ten years from now. Here, we will
discuss the physics case for these experiments based on present--day
knowledge. This is interesting not only because many of the
measurements that we will discuss would only be possible at a neutrino
factory  (for instance, those related to polarized parton
distributions), but also because comparison with what we already know
will allow us to highlight the peculiar features of physics with
neutrino beams.

This brief review is largely based on a recent detailed quantitative
study performed by a CERN working group~\cite{cernrep}; quantitative
estimates given here are taken from there unless otherwise stated. Previous
studies on the physics
potential of neutrino factories have been performed by working groups at
Fermilab~\cite{fermirep} and Brookhaven~\cite{bnlrep}. General
background on neutrino experiments is in Ref.~\cite{nurev}.
Even though we
will usually describe the energy and luminosity dependence of our
results, we will assume the `CERN  reference scenario'~\cite{cernmunu}: 
specifically, a
 50~GeV $\mu$ beam, with $10^{20}$ muon decays per year along a 100~m
straight section.

\section{Neutrino interactions with matter}

The most interesting neutrino--induced reactions in matter are
neutrino--elec\-tron elastic scattering, and neutrino--nucleon
(deep)--inelastic scattering (DIS), i.e. neu\-tri\-no--quark scattering. The
former is a clean purely weak interaction process, while the latter allows
one to probe strongly--interacting matter with weak currents. 

The cross section for elastic neutrino-- or antineutrino--electron scattering 
is
\begin{equation}
\label{elxsec}
\frac{d\sigma}{dy}=\frac{2G_F^2m_eE_\nu}{\pi}
                                    \left[
                                    g_L^2+g_R^2(1-y)^2\right];
\end{equation}
where  $y\equiv
E_e/E_\nu$, $0\le y\le1$, so the total cross section is obtained
replacing $(1-y)^2\to\frac{1}{3}$.
The couplings for neutral--current (NC) processes are
$g_L=\half(g_V+\lambda_\nu
g_A)$,  $g_R= \half(g_V-\lambda_\nu g_A)$, 
with $\lambda_\nu=-1$ ($\lambda_{\bar\nu}=1$), and
for charged--current (CC) processes $g_L=1$ ($g_L=0$), $g_R=0$
($g_R=1$) for neutrinos (antineutrinos). The total couplings are 
listed in the following table (the numerical values are
computed with $\stw=0.23$):
\begin{center}
\begin{tabular}{|c|c|c|c|}\hline
Reaction & $g_L$ & $g_R$ & $g_L^2+\frac{1}{3}g_R^2$ \\ \hline
$\numu e^-\to\numu e^-$ (NC) & $-\frac{1}{2}+\stw$ & $\stw$ & 0.091 \\ 
$\nubmu e^-\to\nubmu e^-$ (NC) & $\stw$ & $-\frac{1}{2}+\stw$ & 0.077 \\ 
$\nue e^-\to\nue e^-$ (NC+CC) & $\frac{1}{2}+\stw$ & $\stw$ & 0.551 \\ 
$\nube e^-\to\nube e^-$ (NC+CC)& $\stw$ & $\frac  {1}{2}+\stw$ & 0.231 \\ \hline
\end{tabular}
\end{center}

The total cross--section is tiny, of order $\sigma\sim
10^{-3}\times (g^2_L+\frac{1}{3} g^2_R)$~pb for 50~GeV neutrinos. Yet  at a
neutrino factory with a 20 ton liquid argon TPC or a fully active 2
ton liquid methane target one expects integrated 
luminosities of order of 
$8.6 \cdot 10^{10} {\rm pb}^{-1}$, leading to  rates
of order of $\sim10^7$ events per year with a $\mu^+$ beam (half with
a $\mu^-$ beam).

The  neu\-tri\-no--nucleon DIS cross section is by a factor
$\sim m_p/m_e$ larger: for charged--current interactions, up to
corrections suppressed by powers of $m_p^2/Q^2$
\bea
\label{disxsec}
&&
\frac{d^2\sigma^{\lambda_p\lambda_\ell}(x,y,Q^2)}{dx dy}
=
\frac{G^2_F}{  2\pi (1+Q^2/m_W^2)^2}
\frac{Q^2}{ xy}\Bigg\{
\left[-\lambda_\ell\, y \left(1-\frac{y}{2}\right) x { F_3(x,Q^2)}
      \right.\nonumber\\ &&\quad\left.
+(1-y) { F_2(x,Q^2)} + y^2 x {
F_1(x,Q^2)}\right]
 -2\lambda_p
  \left[
     -\lambda_\ell\, y (2-y)  x { g_1(x,Q^2)}
\right.\nonumber\\ &&\qquad\left. -(1-y) {
g_4(x,Q^2)}- y^2 x { g_5(x,Q^2)}
  \right]
\Bigg\},
\eea
where $\lambda$ are the lepton and proton helicities (assuming
longitudinal proton polarization), and the
kinematic variables are $y=\frac{p\cdot q}{p\cdot
k}$ (lepton fractional energy loss), $x= \frac{Q^2}{2 p\cdot
q}$ (Bjorken $x$). 
The neutral--current cross--section is found from eq.~(\ref{disxsec})
by letting
$m_W\to m_Z$ and multiplying by an
overall factor  $[\half(g_V-\lambda_\ell
g_A)]^2$.

At a neutrino factory, structure functions could be measured for
$0.01\le x\le 1$ and $1\le Q^2\le100$~GeV$^2$ (note the kinematic
limit $s\equiv \frac{Q^2}{xy}\le 2 m_pE_\mu=$~100~GeV$^2$). Because
$y=Q^2/(2 x m_p E_\nu)$, at fixed $x$ and $Q^2$, $y$ 
only varies with the neutrino energy. 
At a neutrino
factory with a broad--band beam,  neutrinos of various energies
(measurable on an event--by--event basis) are
available.
It is then possible to disentangle all structure functions $F_i$
(unpolarized) or $g_i$ (polarized) by fitting
the $y$ dependence of the data for fixed $x$ and $Q^2$. One expects
statistical errors of order of $1\%$ or better on all three structure
functions for $x\gsim 0.1$ when  $Q^2\lsim 15$~GeV$^2$, and 
for $x\gsim 0.3$ when  $Q^2\lsim 80$~GeV$^2$, and of order $10$\% otherwise.

Information on the structure of the nucleon target is encoded in the
structure functions, whose leading parton content 
in terms of the
unpolarized and polarized quark distribution for the $i$--th flavor
 $q_i\equiv
q_i^{\uparrow\uparrow}+q_i^{\uparrow\downarrow}$ and 
$\Delta q_i\equiv 
q_i^{\uparrow\uparrow}-q_i^{\uparrow\downarrow}$
is summarized in the
following table, where for comparison we also give the standard result
for charged--lepton scattering via virtual photon exchange:
\begin{center}
\begin{tabular}[c]{ccc}

NC&$F_1^{\gamma} =\half\sum_{i}  e^2_i\left(q_i+\bar
q_i\right)$\quad\qquad&$ g_1^{\gamma}=\half\sum_{i}
e^2_i\left(\Delta q_i+\Delta \bar
q_i\right)$\\
NC&$F_1^{Z} =\half\sum_{i}  (g^2_V+g^2_A)_i\left(q_i+\bar
q_i\right)$\quad\qquad&$ g_1^{Z}=\half\sum_{i} (g^2_V+g^2_A)_i
\left(\Delta q_i+\Delta \bar
q_i\right)$\\
NC&$F_3^{Z} =2\sum_{i}  (g_Vg_A)_i \left(q_i+\bar
q_i\right)$\quad\qquad&$ g_1^{Z}=-\sum_{i}(g_Vg_A)_i
\left(\Delta q_i+\Delta \bar
q_i\right)$\\
CC&{ $F_1^\wup =\bar u + d + s + \bar c$}\quad\qquad&${ g_1^\wup=\Delta\bar u + \Delta d +
\Delta s + \Delta \bar c}$\\
CC&${ -F_3^\wup/2 = \bar u - d - s +\bar c }$\quad\qquad & { $g_5^\wup = \Delta \bar u -\Delta d -\Delta s +\Delta\bar c$}\\
\phantom{CC}& $F_2= 2 x F_1$& $g_4= 2 x g_5$\\
\end{tabular}
\end{center}

Here $e_i$ are the electric charges and $(g_V)_i$, $(g_A)_i$ are the weak
charges of the $i$--th quark flavour.
If $W^+\to W^-$ (incoming $\bar \nu$ beam), then
$u\leftrightarrow d,\>c\leftrightarrow s$. The structure functions
$F_3$, $g_4$ and $g_5$ are parity--violating, and therefore not
accessible in virtual photon scattering. 
Of course, beyond leading order in the strong coupling each
quark or antiquark flavor's contribution receives $O(\as)$
corrections proportional to itself and to all other quark, antiquark
and gluon distributions. This last correction is flavor--blind, and
thus decouples from the parity--violating structure functions $F_3$,
$g_4$ and $g_5$. 

\section{Tests of the Standard Model}
\subsection{The weak mixing angle}
The weak mixing angle can be determined from 
the leptonic 
weak couplings $g_L$ and $g_R$, which can be cleanly 
extracted from  the measurement of the elastic cross--section
eq.~(\ref{elxsec}). The main background for this process is
quasielastic neutrino--nucleon scattering. Because the transverse
momentum of the outgoing electron is $p_t\sim\sqrt{m_e E_\nu}$, but
 $p_t\sim\sqrt{m_p E_\nu}$ for scattering off nucleons, the
background can be removed  with a
$p_t$ cut.
The current best determination from this process is 
\begin{equation}
\stw=0.2324\pm0.0058\>{\rm (stat)}\pm0.0059\>{\rm (syst)}
\end{equation}
from the CHARMII experiment~\cite{charmtwo}.

\begin{figure}[h]\vspace{-.2cm}
\begin{center}
\includegraphics[width=0.48\textwidth,clip]{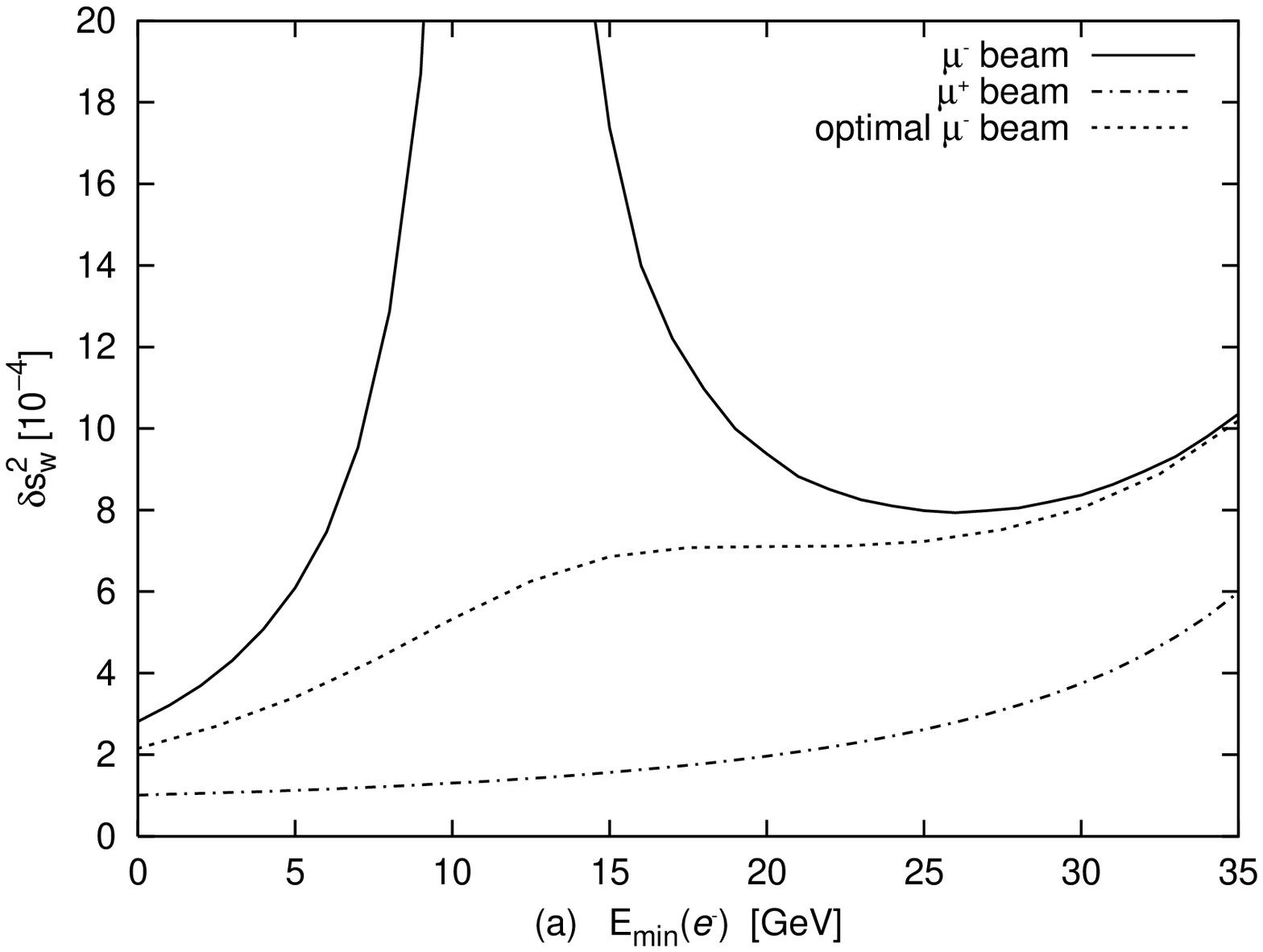} \hfill
\includegraphics[width=0.48\textwidth,clip]{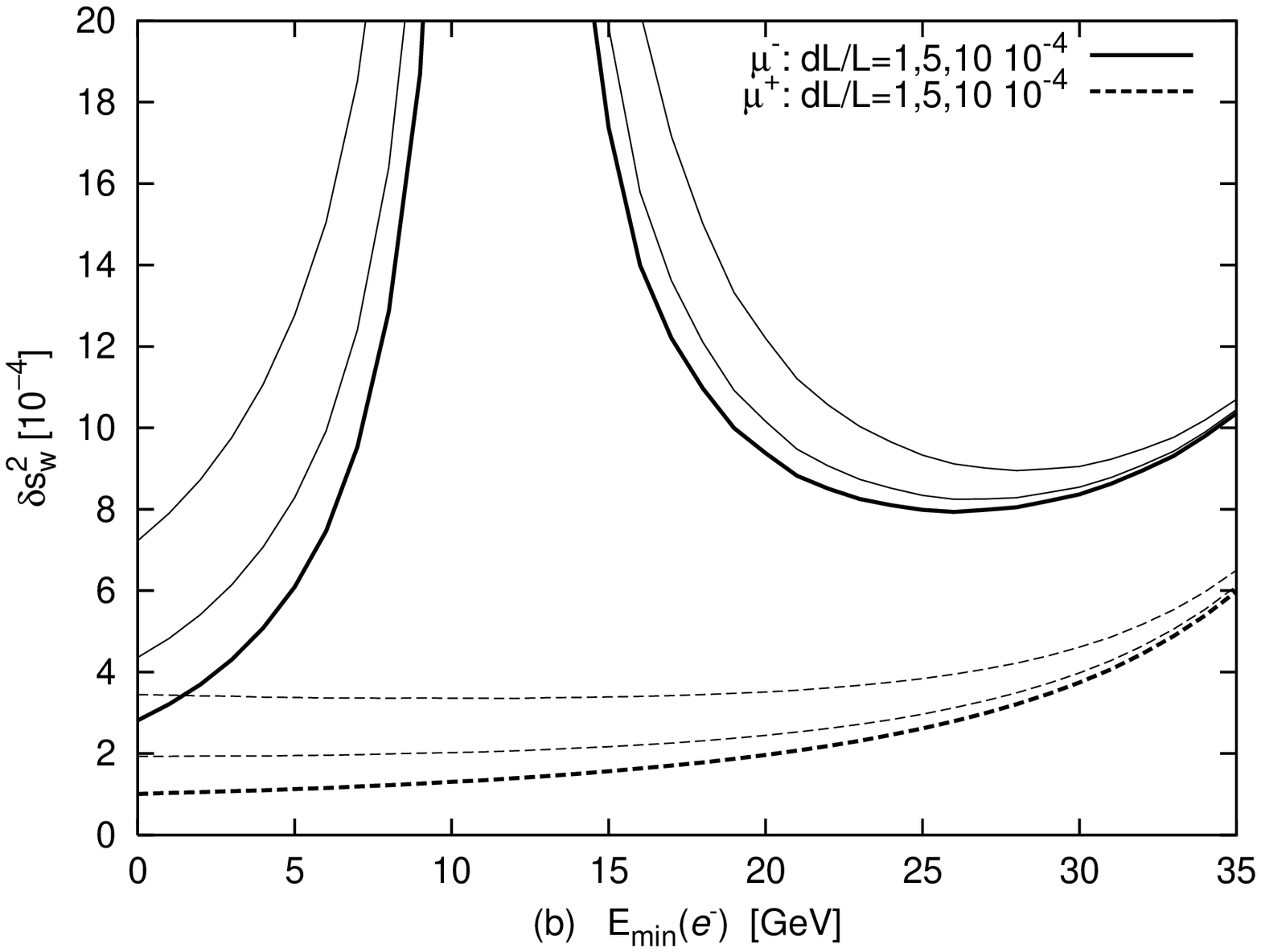} 
\end{center}
\vspace{-.2cm}
\caption{
(a) Statistical uncertainty (in units $10^{-4}$) in the
  extraction of $\sin^2\theta_W$ from $\nu e $ scattering as a
  function of the minimum electron energy.
(b) Impact of luminosity measurement at the level of
  $10^{-3}$, 5 $10^{-4}$, $1\cdot 10^{-4}$ on the same  $\sin^2\theta_W$
sensitivities. 
\label{gambino}}
\end{figure}
At a neutrino factory,
events without a muon in the final state originating from $\nu_\mu$ or
$\bar\nu_e$ (in a $\mu^-$ beam)
cannot be distinguished and must be considered together. 
With the luminosity discussed above, one then
gets the statistical accuracy shown in Fig.~1a. The main systematic
uncertainty is the determination of the incoming neutrino flux (see
Fig.~1b), which can be done by normalizing to the inverse muon decay,
namely the process  $\bar{\nu}_e e^-\to \mu^-\bar{\nu}_\mu$ or 
$\nu_\mu e^-\to \mu^-\nu_e$.

An independent determination of $\stw$ can be obtained from DIS. 
This has the  advantage of larger statistics, but the
disadvantage of systematic uncertainties related to  
knowledge of the nucleon structure.
The value of $\stw$ is extracted from the quark weak couplings:
in  ratios of NC to CC 
DIS cross--sections eq.~(\ref{disxsec}) the lepton couplings
cancel. Also,  such ratios are less sensitive to details of
nucleon structure, in that the leading--order dependence on parton
distributions can  be made to cancel. In particular,
the Paschos--Wolfenstein~\cite{PW} ratio
\beq
\label{pwrel}
R^-=\frac{\sigma_{NC}({\nu}_\mu)-\sigma_{NC}(\bar{\nu}_\mu)}{
\sigma_{CC}({\nu}_\mu)-\sigma_{CC}(\bar{\nu}_\mu)}=\frac12 -
\sin^2\theta_W
\eeq
has been used to obtain a determination~\cite{ccfrstw}
\beq
\!\!\!\!\!\stw({\rm OS})=0.2255\pm 
0.0018 ({\rm stat}) \pm 0.0010 ({\rm syst}).
\eeq
At a neutrino factory, one can only measure combinations where
$\nu_\mu$ and $\bar\nu_e$ do not have to be disentangled on an
event--by--event basis, such as
$R^{\mu^-}=\frac{\sigma_{NC}(\nu_\mu) + \sigma_{NC}(\bar{\nu}_e)}{
\sigma_{CC}(\nu_\mu) + \sigma_{CC}(\bar{\nu}_e)}$, 
$R^{\mu^+}=\frac{\sigma_{NC}(\bar{\nu}_\mu) + \sigma_{NC}({\nu}_e)}{
\sigma_{CC}(\bar{\nu}_\mu) + \sigma_{CC}({\nu}_e)}$, or
$P= \frac{\sigma_{NC}(\mu^-) - \sigma_{NC}(\mu^+)}{
\sigma_{CC}(\mu^-) - \sigma_{CC}(\mu^+)}$, or linear combination
thereof. These combinations give a good handle on $\stw$, but are more
dependent on nucleon structure: in fact, it is convenient to 
construct an optimal combination, which
maximizes the dependence on $\stw$ while minimizing the uncertainty
due to parton distributions.  The uncertainties on
$\stw$ in units of $10^{-4}$ for various combinations are given in the
following table:
\begin{center}
\begin{tabular}{|c|c|c|} 
\hline 
observable & stat. error & PDF 
 \\ \hline
$R^{\mu^-}$ & 0.4 & $\sim$ 12 
 \\ \hline
$R^{\mu^+}$ & 0.5 & $\sim$ 15 
\\ \hline
$R^{\mu^-}-0.8R^{\mu^+}$  & 2.2 & $\sim 2$ 
\\ \hline
$P$ & 4.9 & $\sim$ 4
\\ \hline 
\end{tabular}
\end{center}

In summary, at a neutrino factory the weak mixing angle could be
measured to an accuracy of about $10^{-4}$, in two different
ways. 
This is comparable to the
 best presently available determination of $\stw$,  namely~\cite{moriondew}
$\stw=0.23098\pm0.00026$ from all available
asymmetry measurements. Furthermore, the two measurements from neutrino
elastic scattering and DIS undergo different radiative corrections
from each other and from the asymmetry measurements. This would therefore be
a very competitive test of the standard model, or its violation.

\subsection{The strong coupling}

The strong coupling can be extracted from DIS data either by
considering combinations which do not depend on parton structure (sum
rules), or by performing global fits where both $\as$ and all parton
distributions are simultaneously determined.
In both cases neutrino beams are superior to conventional charged
lepton beams because of the availability of more independent combinations of
individual parton distributions. 

Specifically, if both proton and neutron targets are available,
one may construct two combinations which only depend on
the total number of up plus down valence quarks (Gross--Lewellyn Smith
sum rule~\cite{GLS}) or up minus down valence quarks (unpolarized Bjorken
sum rule~\cite{BSR}), up to an $\as$--dependent factor which is currently
known to $O(\as^3)$: 
\bea
\label{sr}
&&S_{\rm GLS}^N(Q^2) = \frac{1}{2}\int_0^1 dx\,\left(F_3^{\nu p}(x,Q^2)
+F_3^{{\nu} n}(x,Q^2)\right)\nonumber\\
&&\quad={ C_{\rm GLS}(Q^2)}{ \int_0^1 dx\, \left[u(x,Q^2)-\overline{u}(x,Q^2)+d(x,Q^2)
-\overline{d}(x,Q^2)\right]}\nonumber\\
&&\quad={ 3}{\left(1-\frac{\as(Q^2)}{\pi}-
3.25\left[\frac{\as(Q^2)}{\pi}\right]^2
-12.2\left[\frac{\as(Q^2)}{\pi}\right]^3+\dots
\right) + \frac{h_{\rm GLS}}{Q^2}}\nonumber\\
&&S_{\rm BjU}^N(Q^2) = \frac{1}{2}\int_0^1 dx\,\left(F_1^{\nu p}(x,Q^2)
-F_1^{{\nu} n}(x,Q^2)\right)\\
&&\quad={ C_{\rm BjU}(Q^2)}{\int_0^1 dx\,
\left[u(x,Q^2)-\overline{u}(x,Q^2)-\left(
d(x,Q^2)
-\overline{d}(x,Q^2)\right)\right]}\nonumber\\
&&\quad={ 1}{ \left(
1-\frac{2}{3}\frac{\alpha_s(Q^2)}{\pi}-
2.65\left[\frac{\as(Q^2)}{\pi}\right]^2
-13.38\left[\frac{\as(Q^2)}{\pi}\right]^3+\dots\right)+\frac{h_{\rm
BjU}}{Q^2}},
\nonumber
\eea
where the coefficients of the power corrections 
$h_{\rm GLS}$ and $h_{\rm
BjU}$ are unknown and must be fitted.
The disadvantage of this way of determining $\as$ is the need to
extrapolate over the full range $0\le x\le1$ the data which are only
available in a limited range of $x$. The accessible $x$ range is
larger at lower $Q^2$, where however  power suppressed
corrections are larger. 
At present, a determination of $\as(M_Z)$ with an error 
$\Delta \alpha_s(M_Z)=\epm{0.009}{0.012}$ has been obtained from the GLS
sum rule with the CCFR neutrino beam~\cite{ccfrgls}, while
the BjU  integral has never been measured because of the
impossibility of disentangling $F_1$ and $F_2$ from present--day neutrino
DIS data.  At the neutrino factory, one could
reach an uncertainty $\Delta \alpha_s(M_Z)=0.0035$, which is completely
dominated by  limited kinematic coverage in either $Q^2$ (power
corrections) or $x$ (small $x$ extrapolation) 
and could only be improved if a higher
energy beam were available.

A very competitive determination could be obtained by performing a
global fit to structure functions: interestingly, because more
structure functions are available, at a neutrino
factory the full set of parton distributions could be determined in a
single experiment (see Sect.~4 below). 
One could then achieve a statistical accuracy on $\as$ of
order $\Delta \alpha_s(M_Z)=0.0003$, 
 to be compared to the statistical accuracy 
$\Delta \alpha_s(M_Z)\approx0.0017$ which can be obtained from present--day
global
fits~\cite{alek} which include both $\nu$ and charged--lepton DIS data from
various experiments (the error from current neutrino experiments alone
is of order $\Delta \alpha_s(M_Z)\approx0.005$~\cite{alkat}). The
statistical precision of  this determination of $\as$  
value is by one order of magnitude better than 
the extant  global error~\cite{albet}
$\Delta\alpha_s(M_Z)\sim 0.003$. The accuracy in the determination of
$\as$ at a neutrino factory would thus be entirely dominated by 
theoretical uncertainties,  
and it could be the most accurate determination once
next-to-next-to-leading order corrections to perturbative evolution~\cite{ho}
are known. 

More studies of strong interaction physics could take advantage of
the fact that charm is
copiously produces in charged--current events and easily
detected. This could be exploited not only in order to further refine
our knowledge of parton distributions through the study of specific
semi--inclusive channels, but also to study QCD corrections
to charm production near threshold, and finally to  measure accurately
absolute branching ratios and decay constants of individual
charmed mesons, such as $\Lambda_c$ or $D_s$.

\section{The Structure of the Nucleon}
Current information on the parton distributions of the nucleon~\cite{pdfs}
comes mostly from
DIS data, the bulk of which are produced with charged lepton
beams and thus are essentially NC scattering. The problem is then that
(see Sect.~2) only one combination of $q_i+\bar q_i$ distributions
is accessible. This means that different quark flavors can only be
disentangled using isospin, if proton and neutron targets are
available, plus in principle by exploiting perturbative evolution (i.e. 
subleading corrections). Hence, using NC DIS data it is hard
to determine strangeness, and impossible to measure the C--odd
combination $q-\bar q$.

This is to be contrasted with the situation in CC scattering~\cite{RDBnu},
where if both
proton and neutron targets and $\nu$ and $\bar \nu$ beams are
available, then one can form eight  linear combinations of
the two independent structure functions
($F_1$ and $F_3$ unpolarized, $g_1$
and $g_5$ polarized), six of which are independent (NC data do not
give any extra independent information).  
It is easy to construct leading--order combinations
of parton distributions which, below charm threshold, determine all
six light flavours and anti--flavours independently. The charm
distribution can then be determined either comparing data
below and above charm threshold, or tagging charm events,
which have a distinct dimuon signature~\cite{nurev}.
In practice, of course, parton distributions will be determined by fitting
the full next--to--leading order expression of structure functions,
however  the fact that individual partons can already be disentangled 
at a leading--order guarantees the accuracy of  the NLO determination.

\subsection{Unpolarized DIS: the flavor content of the nucleon}

\begin{figure}[h]\vspace{-.2cm}
\begin{center}
\includegraphics[width=0.47\textwidth,clip]{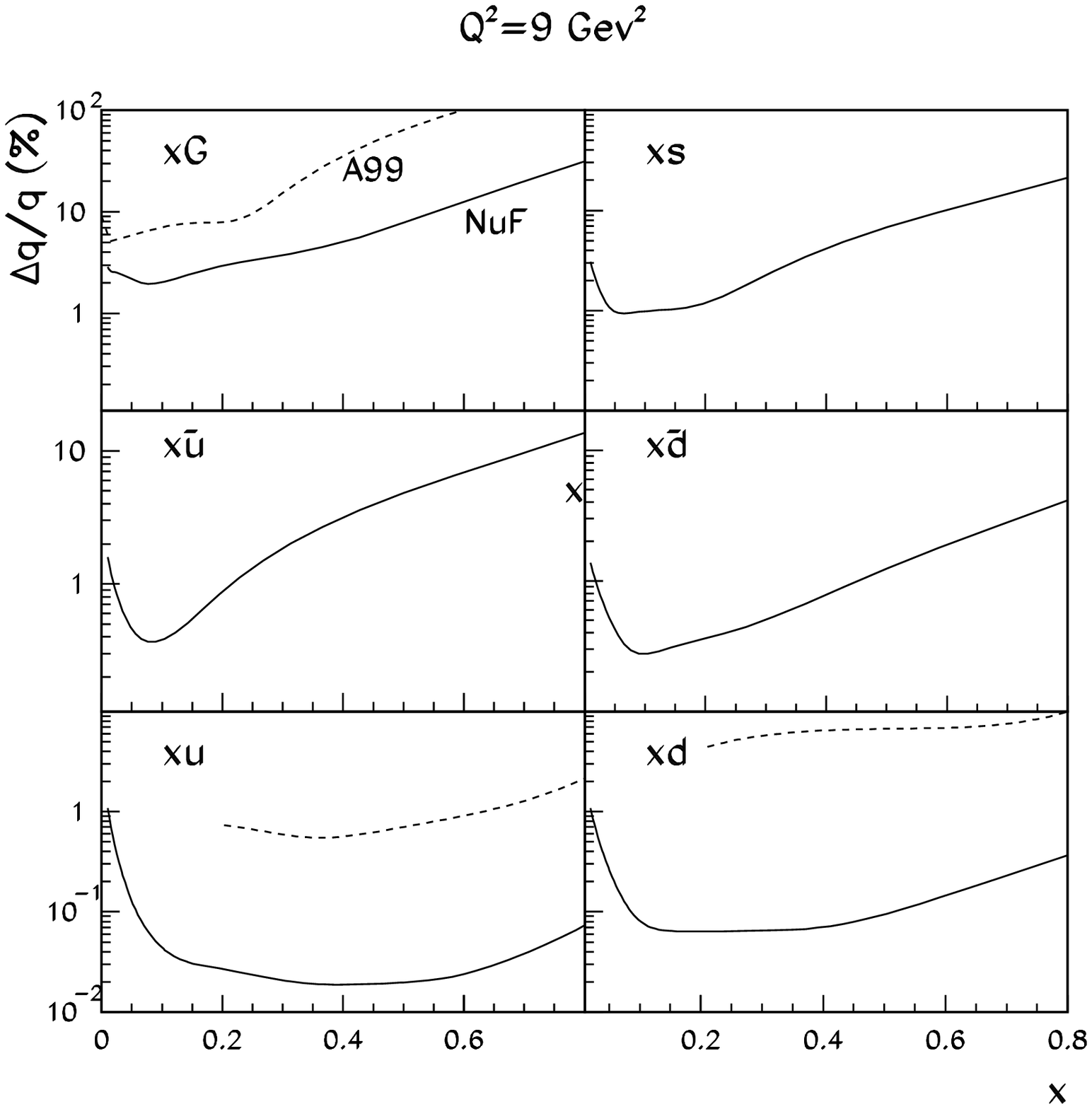} \hfill
\includegraphics[width=0.47\textwidth,clip]{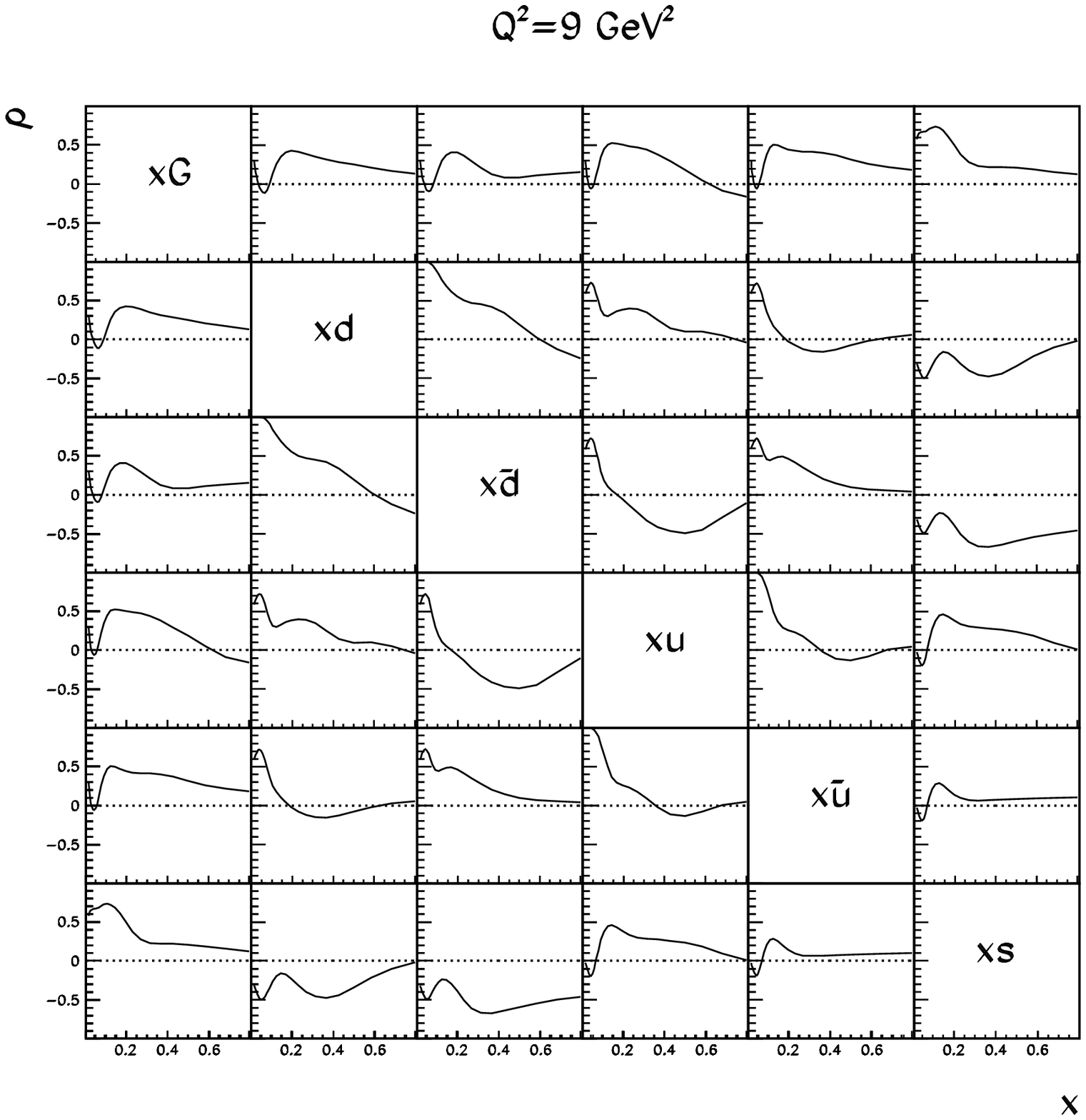} 
\end{center}
\vspace{-.2cm}
\caption{
(a) Percentage error on parton distributions at a neutrino factory
(solid) compared with 
present--day~\cite{alek} errors (dashed).
(b) Correlation coefficients between parton distributions determined
at a neutrino factory.
\label{alekhin}}
\end{figure}

Unpolarized parton distributions are a necessary ingredient in the
computation of any collider process. However, only the up, down and
gluon distributions can be determined in a reasonably accurate way
from DIS data.
Some information on strangeness can be extracted~\cite{BPZ} 
from neutrino data, while some less--inclusive observables (such as
$W$ production, or Drell--Yan)  provide some constraints on the
relative size of the $q$ and $\bar q$ distributions, but the results
are at best semi--quantitative (see Fig.~2a).

A quantitative estimate of the
accuracy at a neutrino factory can be obtained by generating
pseudo--data for structure functions 
with appropriate errors, and then producing a fit of parton
distributions based on this data.
In Fig.~2a the error estimates on individual partons obtained in this
way are compared to the extant knowledge. 
Note that no current errors on strange and
antiquark distributions are given, since the present results largely depend
on theoretical prejudice.
 In Fig.~2b we further show that the
point--by--point correlation of individual distributions is uniformly
quite low, indicating that a model--independent
flavour and antiflavour separation is possible to 10\%--20\% accuracy
in most of the accessible kinematic range.
A  neutrino factory would essentially turn the
determination of individual parton distributions into a
precision quantitative exercise. 

\subsection{Polarized DIS: the spin of the nucleon}

Polarized DIS has recently attracted considerable attention because 
the unexpected smallness of the proton's singlet axial charge $a_0$
suggests that
the nucleon spin structure is considerably subtler than naive parton
expectations might suggest~\cite{spinrev}. In the naive parton
model the singlet axial charge is the fraction of the nucleon spin
which is carried by quarks; the Zweig rule would lead one to expect
this to be around 60\% but the experimental value is compatible with
zero. 

This state of affairs point to several possible scenarios for
the nucleon spin structure. 
Beyond leading order the axial charge is given by
\beq\label{anom}
a_0= \Delta \Sigma-\frac{n_f\as}{2\pi} \Delta G,
\eeq
 where $\Delta
\Sigma=\sum_i(\Delta q_i+\Delta \bar q_i)$ is the scale--invariant
quark spin fraction, and $\Delta G$ is
the gluon spin fraction. The latter, due to the axial anomaly,  gives an
effectively leading--order contribution to $a_0$~\ref{anom} (it depends on
scale as $\Delta G\sim \frac{1}{\as}$). 
So a first possibility is that $\Delta G$ is large enough that the
quark spin $\Delta
\Sigma$ is large even though the axial charge $a_0$ is small
(`anomaly' scenario~\cite{anom}). A  
different  option (`instanton' scenario) is that  $\Delta\Sigma$ is itself small, because
of a large contribution from sea
quarks whose polarization is anticorrelated to that of valence quarks
(possibly because of `instanton' QCD vacuum configurations~\cite{inst}). Yet
another possibility (`skyrmion' scenario) is that $\Delta\Sigma$ is small because of a
large contribution from `valence' strange quarks $|\Delta s|
>>|\Delta\bar s|$ (as suggested~\cite{skyr} in the
Skyrme model).

 At present, the quark and gluon spin fractions can be
extracted from NC DIS data~\cite{abfr}: 
$\Delta G(1,1\,{\rm GeV}^2)=
0.8\pm0.2$,
$\Delta \Sigma (1) = 0.38\pm0.03$, while the  `octet' combination can
be determined using SU(3) from baryon $\beta$--decay constants: $a_8\equiv\Delta u+\Delta
d-2\Delta s=0.6\pm 30\%$ (the large error comes from a conservative
estimate of SU(3) violation). It is of course impossible to polarize
the kind of targets which are required for present--day neutrino DIS
experiments, so no neutrino data are available, and thus  little
information on strangeness and no information at all 
on the quark--antiquark separation is available in the polarized case.
At a neutrino factory, significant rates could be achieved with small
targets~\cite{RDBnu}: with a detector radius of 50~cm, 100~m length, 
the structure functions $g_1$,  $g_5$ could be
independently measured to an accuracy which is about one order of
magnitude better than that with which $g_1$ is determined 
in present charged lepton DIS experiments.   

We can then assess the
information that one might obtain on the nucleon spin structure by
generating pseudodata within different representative
scenarios~\cite{fmr}. 
Assuming an `anomaly' scenario one would determine
$\Delta
g=0.9\pm0.1$; $\Delta \Sigma  = 0.39\pm0.01$; $a_8=0.56\pm 0.01$,
while in an `instanton'  scenario 
$\Delta g=0.2\pm0.1$; $\Delta \Sigma  = 0.32\pm0.01$;
$a_8=0.57\pm 0.01$. Furthermore, in an `instanton' scenario
 $\left[\Delta s-\Delta\bar s\right](1,1\,{\rm
GeV}^2)=-0.007\pm0.007$; while in a 
`skyrmion''  scenario one would observe $\left[\Delta s-\Delta\bar s\right](1,1\,{\rm
GeV}^2)=-0.106\pm0.008$. Note that the strange contribution, and
hence the octet component, would be determined from the data directly,
without having to use SU(3) (which would thus be tested in the process).
Clearly, the distinct scenarios could be well separated from each
other: in fact
 full flavor separation 
at the level of first moments would be possible.  Hence, a 
resolution of the nucleon spin structure would be possible at a
neutrino factory.

The situation would be less favorable for the $x$--dependence of
parton distributions, because of the difficulty of disentangling the
potentially large gluon contribution. If the gluon were well
determined from other experiments, however, a full point--by--point
determination of the polarized parton content of the nucleon would be
possible, 
to an accuracy comparable to that
of the unpolarized case.
\section{Exotica}

On top of the standard processes discussed so far, the high luminosity
at a neutrino factory allows the study of several rare or exotic
processes. of which we give two examples.

\begin{figure}[h]\vspace{-.2cm}
\begin{center}
\includegraphics[width=0.6\textwidth,clip]{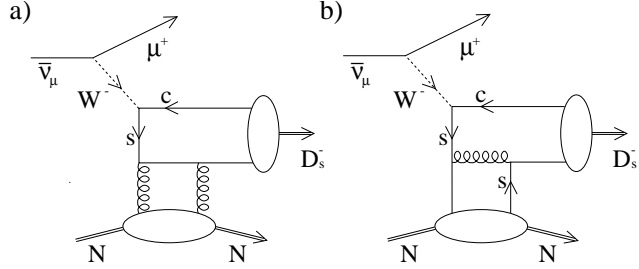} 
\end{center}
\vspace{-.2cm}
\caption{
Feynman diagram for hard exclusive $D_s$ production
\label{SPD}}
\end{figure}
A first example is the extraction of off-diagonal (or skewed) parton
distributions (SPD)  from hard exclusive meson production. The Feynman
diagrams for this process are shown in Fig.~3. If the virtuality $Q^2$
of the $W$ boson  is  large compared to the nucleon momentum transfer
$t$ and all masses, then the cross section for
the process factorizes~\cite{spdfac} as
$ \frac{d\sigma}{d x_{Bj} d Q^2 d t}(W^-+N\to D^-_s+N)=H\otimes
\Phi_D\otimes F$, where
$H$ is the cross--section for the underlying
hard  perturbative parton subprocess,
$\Phi_D$ is the  $D^-$ fragmentation function, and
$F(x,t,Q^2)$ is an SPD,
which interpolates between the usual nucleon form factor $G(t)$ (which
is related to the first $x$--moment of $F$) and parton distribution
$F(x)$ (related to the $t\to0$ limit of $F$). The estimated cross
section for this process 
is $\sigma=2.2\times
10^{-5}$~pb, leading to  $10^4$ event/yr at the neutrino factory. 
Given good knowledge of the $D_s$ fragmentation, the 
SPD can   be measured. Because of
relatively low backgrounds and the distinct experimental signature,
this would be a favourable way of measuring SPDs, which have 
never been determined so far.

As another interesting `exotic' process, consider neutrino--electron
annihilation into hadrons through radiation of an intermediate virtual
$W$~\cite{bnlrep}. 
One can then define 
the analogue of $R$-ratio of $e^+e-$ annihilation:
\beq
\label{rrat}
 R_A\equiv\frac{\sigma( \bar\nu_{e}e^{-}\rightarrow {\rm hadrons})}{
\sigma(\bar\nu_{e}e^{-}\rightarrow\bar\nu_{\mu}\mu^{-})}.
\eeq
$R_A(s)$ is the spectral function for annihilation into
axial vector final states:
at large  $\sqrt{s}$, $R_A(s)$ can be computed in 
perturbative QCD from the underlying parton processes, while
at low $\sqrt{s}\approx m_\pi$,   $R_A(s)$ can be determined
using PCAC from the relevant axial current matrix element.
The threshold energy for the inverse muon decay process
($\bar\nu_{\mu}\mu^{-}$ final state) is $E_\nu=11$~GeV, while the
threshold for the lowest hadronic final state $\gamma \pi^-$ is
$E_\nu=19$~GeV, so at a neutrino factory one could perform some
interesting tests of PCAC and chiral symmetry.
\section{New Physics}

All the standard physics items which we discussed so far can also be
viewed as tests for new physics, or tools in searches for new
physics. Specifically, different determinations of the weak mixing angle
lead to tests of the electroweak sector through the comparison of the
corresponding radiative corrections (Sect.~3.1). Determinations of
the  strong coupling (Sect.~3.2) and especially its running test 
the strongly interacting sector. Precision
determinations of parton distributions (Sect.~4) are a necessary input
in searches for new physics: in fact some recent possible indications
of new physics could also be explained by invoking lack of accurate knowledge
of parton distributions~\cite{ctequnc}.

However, high--intensity neutrino beams can also be used to design
specific searches for new physics, of which we also give two
examples. 
First, one can exploit the
copious production of charmed mesons to study specific non--standard
decay channels. An interesting possibility is the search for possible
T--violation in the decays of the $\Lambda_c$~\cite{bnlrep}. Indeed,
in the semileptonic decay $\Lambda_c^+\to l^+\nu\Lambda$
one can construct an experimentally observable T--odd correlation
$C_{T} \equiv 
\langle \vec \sigma _{\Lambda} \cdot ( \vec p_{\Lambda} 
\times \vec p_l) \rangle$ from the spin and momentum of the daughter
hyperon and the momentum of the
final--state lepton . This correlation cannot be
affected by either strong or electromagnetic final state interactions,
so a nonzero value for it is a measure of T--violation.

Another example is the search for lepton--flavour violating decays,
such as  $\mu\to e \bar\nu_\tau\nu_i$. The corresponding four--Fermi
couplings  could be probed at a neutrino factory down to a strength
about  $\sim 3\times 10^{-4}$ weaker than ordinary weak interactions. 
Of course stronger bounds already exist in the
charged lepton sector, but the current limits for processes involving
neutrinos
are  only of order
$\sim 10^{-1}-10^{-2}$.
 
\section{Outlook}
The peculiar features of a neutrino beam is the availability of a
probe which depends both on spin and flavor.
However, the physics potential of such a beam can only be exploited
given high enough intensity. Consequently, there is a whole class of
measurements which is only possible at a neutrino factory. An
example is charged--current polarized deep--inelastic scattering,
which can only be studied accurately with a neutrino beam and a
reasonably sized target: with charged lepton beams it is hard to
separate charged--current and neutral--current events, and with low
intensities it is impossible to polarize the target.
Which of these measurements might be the most interesting at a
neutrino factory will largely depend on the development of
high--energy physics in the next decade. In any case, a neutrino
factory has the potential of developing a broad program of
short--baseline physics, which could answer many outstanding questions
which are not likely to be addressed at any other experimental
facility.\\
{\bf Acknowledgements}: I thank Y.~Kuno for inviting me to this
very stimulating meeting, A.~Donini for several 
discussions during the conference, and R.~Ball and G.~Ridolfi for a
critical reading of the manuscript.

\end{document}